\documentclass[final]{IEEEtran}
\usepackage{amsthm,amssymb,graphicx,multirow,amsmath,color,amsfonts}
\usepackage[update,prepend]{epstopdf}
\usepackage[noadjust]{cite}
\usepackage[latin1, utf8]{inputenc}
\usepackage{tikz}
\usepackage{bbm} 
\usepackage{pdfpages}
\usepackage{tabulary}
\usepackage{multirow}
\usepackage{comment}
\usepackage{subfigure}
\usepackage{float}
\usepackage[titlenumbered,ruled]{algorithm2e}
\usepackage[noend]{algpseudocode}
\DeclareUnicodeCharacter{0177}{\^y}

\def\nb0{{\mathbf{0}}}
\def\nb1{{\mathbf{1}}}

\newtheorem{lemma}{Lemma}

\newtheorem{definition}{Definition}

\newtheorem{theorem}{Theorem}

\newtheorem{remark}{Remark}

\allowdisplaybreaks 
\begin{document}
\title{
	A Dominant Interferer-based Approximation for Uplink SINR Meta Distribution \\
	in Cellular Networks
}
\author{
	Yujie Qin, Mustafa A. Kishk, {\em Member, IEEE}, and Mohamed-Slim Alouini, {\em Fellow, IEEE}
	\thanks{Yujie Qin and Mohamed-Slim Alouini are with Computer, Electrical and Mathematical Sciences and Engineering (CEMSE) Division, King Abdullah University of Science and Technology (KAUST), Thuwal, 23955-6900, Saudi 
		Arabia. Mustafa Kishk is with the Department of Electronic Engineering, Maynooth University, Maynooth, W23 F2H6, Ireland. (e-mail: yujie.qin@kaust.edu.sa; mustafa.kishk@mu.ie; slim.alouini@kaust.edu.sa).} 
	
}
\date{\today}
\maketitle
\begin{abstract}
This work studies the signal-to-interference-plus-noise-ratio (SINR)  meta distribution for the uplink transmission of a Poisson network with Rayleigh fading by using the dominant interferer-based approximation. The proposed approach relies on computing the mix of exact and mean-field analysis of interference. In particular, it requires the distance distribution of the nearest interferer and the conditional average of the rest of the interference. Using the widely studied fractional path-loss inversion power control and modeling the spatial locations of base stations (BSs) by a Poisson point process (PPP), we obtain the meta distribution based on the proposed method and compare it with the traditional  beta approximation, as well as the exact results obtained via Monte-Carlo simulations. Our numerical results validate that the proposed method shows good matching and is time competitive.  
\end{abstract}

\begin{IEEEkeywords}
	Meta distribution, approximation, stochastic geometry, Poisson network, reliability, uplink
\end{IEEEkeywords}

\section{Introduction}

For a wireless network, the system's performance depends heavily on the spatial configuration of nodes. The accurate modeling of the locations of nodes provides key system insights on the network performance \cite{elsawy2013stochastic}. Stochastic geometry provides tractable mathematical tools to capture the randomness of the nodes \cite{elsawy2016modeling}. While investigating the spatial average performance metrics, such as coverage probability, is important, the performance of an individual link can no longer be ignored, as it reveals the reliability and quality of service of a network \cite{haenggi2015meta}. On the side of operators, obtaining information about the percentile of users achieving a certain performance measure is a fundamental design objective.
This high-level system insight is so-called SINR/SIR meta distribution, defined as a complementary cumulative distribution function (CCDF) of the conditional success probability (probability that the user equipment (UE) achieves cellular coverage, conditioned on the realization of point processes), and used to investigate the discrepancies among UE \cite{elsawy2017meta}.

 The exact expression of SINR/SIR meta distribution is typically computed by using Gil-Pelaez theorem \cite{gil1951note}, however, it is generally hard to obtain as it requires imaginary moments. Some inequalities, such as Markov, Chebyshev, and Chernoff, were provided in \cite{haenggi2015meta} to bound the meta distribution. A comprehensive introduction about meta distribution, as well as some examples on PPP networks and Poisson bipolar networks, was provided in \cite{haenggi2021meta1,haenggi2021meta2}. Specifically, in \cite{haenggi2021meta2}, the author provided a closed-form result of SIR meta distribution for PPP and Poisson bipolar network by using nearest-interferer-only approximation. Authors in \cite{9072358} computed the SIR meta distribution for any independent fading and analyzed the separability. Authors in \cite{ganti2015asymptotics} approximated the meta distribution of a general network by shifting the meta distribution of a Poisson network. 
 
For investigating SIR/SINR meta distribution in different scenarios, authors in \cite{9462466} studied the Poisson line Cox bipolar networks and Poisson stick line Cox bipolar networks; authors in \cite{cui2017sir} analyzed cellular networks with BS cooperation; authors in \cite{saha2020meta} characterized a $K$-tier HetNet combined by PPP and MCP; and authors in \cite{mankar2019meta} considered a typical cell. Besides, the SINR meta distribution of UAV-involved networks was studied in \cite{10024838,qin2023downlink}.

In our previous work \cite{10066317}, we proposed a dominant interferer-based approximation as an alternative method of computing downlink meta distribution in Poisson networks with Rayleigh fading, and it shows good matching with the exact value. In this work, we extend it into uplink scenarios and complete the analysis of the proposed approximation. Different from downlink scenarios, the system models of the uplink case are more complex, for instance, the locations of interferers are modeled by a  non-homogeneous PPP, the nearest interferer may be closer than the typical UE to the BS, and the transmit power is a function of the distance to the serving BS. 


\section{System Model}
This work focuses on the uplink transmission of a PPP wireless cellular network. The locations of BSs are distributed according to a PPP, denoted by $\Phi$ with density $\lambda$. Assuming BSs split the resources to serve the UE, therefore, the UE point process is generated by independently and uniformly selecting one UE in each Voronoi cell formed by $\Phi$. From the perspective of a BS, the locations of interferers can be approximately modeled as a non-homogeneous PPP, $\Phi_i$, with intensity function $\lambda_i(d) = \lambda(1-\exp(-\pi\lambda d^2))$, where $d$ denotes the distance to the BS \cite{andrews2016primer}.
Here, we condition a BS to be at the origin and this BS becomes the typical BS when averaging over the point process. We focus on the SINR of the typical BS which is equivalent to any other arbitrary deterministic location owing to the stationarity of PPP. 

When UE is served by its nearest BS, a fractional path-loss inversion power control with compensation factor $\epsilon$ and a standard path-loss model with exponent $\alpha>2$ are used. Besides,  a Rayleigh fading channel model is used in this work.

By using the inversion power control technology \cite{6786498,elsawy2017meta}, the truncated transmit power of a user in the uplink transmission is given by
\begin{align}
	p_{t}(R_u) =  \left\{
	\begin{aligned}
	&\rho R_{u}^{\alpha\epsilon},\quad \text{if}\quad 0< R_u < \bigg(\frac{p_{\max}}{\rho}\bigg)^{\frac{1}{\alpha\epsilon}},\\
		&p_{\rm max}, \quad \text{if}\quad R_u \geq \bigg(\frac{p_{\max}}{\rho}\bigg)^{\frac{1}{\alpha\epsilon}},\\
	\end{aligned}
	\right.
\end{align}
where $R_u$ is the Euclidean distance between the user and the BS, $\rho $ is a power control parameter to adjust the received power at the serving BS, $\epsilon $ is the compensation factor, and $p_{\rm max}$ is the maximum transmit power.
Consequently, the received power at the BS is
\begin{align}
p_r = H p_t(R_u)R_{u}^{-\alpha},
\end{align}
in which $H$ present the fading gain which follows exponential distribution with mean of unity.

The uplink SINR of the typical BS is
\begin{align}
	{\rm SINR} = \frac{H p_r}{\sigma^2+I} = \frac{H p_{t}(R_u) R_{u}^{-\alpha}}{\sigma^2+I},
\end{align}
where $\sigma^2$ is the thermal noise and $I$ is the aggregate interference,
\begin{align}
I &= \sum_{x\in\Phi_i} H_x p_{t,x}(R_{i,x}) D_{i,x}^{-\alpha},
\end{align}
in which $R_{i,x}$ is the distance between the interferer $x$ to its serving BS, $H_x$, $p_{r,x}$ and $D_{i,x}$ denote the channel fading, transmit power and distances of the interferer $x$, respectively. In what follows, the conditional success probability of the typical BS is given by
\begin{align}
P_s(\theta) = \mathbb{P}({\rm SINR} > \theta\mid\Phi_i).
\end{align}

For an arbitrary realization of $\Phi_i$, we analyze the SINR meta distribution, which is defined as the fraction of links achieving the uplink coverage (SINR above $\theta$), and its argument denotes the system reliability.
\begin{definition}[SINR Meta Distribution] The SINR meta distribution of the uplink is defined in \cite{haenggi2015meta} as
\begin{align}
\bar{F}_{P_{s}}(\theta,\gamma) = \mathbb{P}(P_{s}(\theta) > \gamma),\label{eq_meta_ext}
\end{align}
where $\gamma\in[0,1]$.
\end{definition}
\section{Mathematical Analysis}
In this section, we provide the analysis for the dominant interfer-based approximation for SINR meta distribution in the case of uplink transmission. To compare the proposed approximation with the traditional method, we also derive the beta approximation, which is based on the first and the second moment of the conditional success probability. To do so, we first derive some important distance  distributions.

Recall that $R_u$ denotes the distance to the serving BS, $R_i$ and $D_i$ are the distances between the interferer to its serving BS and typical BS, respectively, and $D_1$ is the distance from the typical BS to the first nearest UE $D_1 = \min(D_{i,x})$, e.g., $D_1 = D_{i,x_1}$.  
\begin{lemma}[Distance Distribution]\label{lemma_distance}
	The probability density function (PDF) and cumulative distribution function (CDF) of $R_u$ is given by
\begin{align}
f_{R_u}(r) &= 2\pi\lambda\exp(-\lambda\pi r^2), \quad r\geq 0,\nonumber\\
F_{R_u}(r) &= 1-\exp(-\lambda\pi r^2), \quad r\geq 0. \label{eq_F_Ru}
\end{align}
Based on the strongest association police, $R_i$ is upper bounded by $D_i$, which forms a truncated Rayleigh distribution, and given by
\begin{align}
	f_{R_i}(r|D_i) = \frac{2\pi\lambda\exp(-\lambda\pi r^2)}{1-\exp(-\lambda\pi D_{i}^{2})},\quad 0\leq r \leq D_i.
\end{align}
The PDF of $D_1$ is given by
\begin{align}
	f_{D_1}(r) =& 2\pi\lambda(1-\exp(-\pi\lambda r^2))r\nonumber\\
	&\times\exp(-2\pi\lambda\int_{0}^{r}(1-\exp(-\pi\lambda z^2))z{\rm d}z).
\end{align}
\end{lemma} 


In what follows, we compute the SINR meta distribution based on the proposed dominant interferer-based approximation and beta approximation, which are provided in the next two subsections.

\subsection{Proposed Approximation}
As mentioned in \cite{10066317}, the closer interferer has higher impact on the system performance and dominant interferer-based approximation works well in downlink transmission. Similarly, we follow the same method and rewrite the interference term by considering the nearest interfering UE exactly while the rest of the interferers in average. To simplify the notation, we use $R_{i,1}$ and $p_{t,1}(R_{i,1})$ to represent $R_{i,x_1}$ and $p_{t,x_1}(R_{i,x_1})$, and $p_{t}^{\prime}(r) = \min(p_{\rm max},\rho r^{-\alpha\epsilon})$.

\begin{lemma}[Approximated Interference]\label{lemma_approx_int}
The approximation of the interference is
\begin{align}
I_1 \approx H  p_{t,1}(R_{i,1}) R_{1}^{-\alpha} +G(D_{1}),
\end{align}
in which
\begin{align}
	G(D_{1}) 	=& 2\pi\lambda\int_{D_{1}}^{\infty}\int_{0}^{z} p_{t}^{\prime}(r)2\pi\lambda\exp(-\pi\lambda r^2) z^{-\alpha+1} {\rm d}r{\rm d}z.
\end{align}
\end{lemma}
\begin{IEEEproof}
See Appendix \ref{app_approx_int}.
\end{IEEEproof}

Based on the approximated aggregate interference, the approximation of the conditional success probability is given in the following lemma.

\begin{lemma}[Approximated Conditional Success Probability]\label{lemma_Ps_appro}
The conditional success probability is approximated by
\begin{align}
	&\mathbb{P}({\rm SINR} > \theta \mid R_{u})\nonumber\\
	& \approx \exp\bigg(\theta(\sigma^2+G(D_{1})) R_{u}^{\prime -1}\bigg) \frac{1}{1+\theta p_t(D_{1})D_{1}^{-\alpha} R_{u}^{\prime -1}},
\end{align}
in which $R_{u}^{\prime} = p_{t}(R_u) R_{u}^{-\alpha}$.
\end{lemma}
\begin{IEEEproof}
See Appendix \ref{app_Ps_appro}.
\end{IEEEproof}

Now we are able to proceed to the final expression of the proposed approximation in uplink scenarios.

\begin{theorem}[Approximated SINR Meta Distribution]\label{theorem_meta_app}
The SINR meta distribution is approximated by
\begin{align}
	&\bar{F}_{P_s}^{\prime}(\theta,\gamma)  \approx \int_{0}^{\infty}F_{R_{u}^{\prime}}(K(r,\theta,\gamma))f_{D_1}(r){\rm d}r,
\end{align}
in which 
\begin{align}
	K(r,\theta,\gamma) &\approx -\frac{1}{\kappa(r)}+\frac{1}{S(r)}W\bigg(0,\frac{S(r)\exp(S(r) \kappa^{-1}(r))}{\gamma\kappa(r)}\bigg),
\end{align}
where $\kappa(r)= \theta \bar{p}_{t}(r) r^{-\alpha}$ and $S(r) = \theta(\sigma^2+G(r))$,
\begin{align}
	\bar{p}_{t}(r) &= \int_{0}^{r} p_{t}^{\prime}(r)f_{R_i}(x|r){\rm d}x,\nonumber\\
	F_{R_{u}^{\prime}}(x) &= \left\{
	\begin{aligned}
		&F_{R_u}((x\rho)^{\frac{1}{(1-\epsilon)\alpha}}), \quad \text{if}\quad x < p_{\rm max}\bigg(\frac{p_{\rm max}}{\rho}\bigg)^{ \frac{1}{\epsilon\alpha}-\alpha},\nonumber\\
		&F_{R_u}((xp_{\rm max})^{\frac{1}{\alpha}}),\quad \text{if}\quad x > p_{\rm max}\bigg(\frac{p_{\rm max}}{\rho}\bigg)^{ \frac{1}{\epsilon\alpha}-\alpha},\nonumber\\
	\end{aligned}
	\right.
\end{align}
where $F_{R_u}(r)$ is the CDF of $R_u$ given in (\ref{eq_F_Ru}) Lemma \ref{lemma_distance}.
\end{theorem}
\begin{IEEEproof}
See Appendix \ref{app_theorem_meta_app}.
\end{IEEEproof}

\subsection{Beta Approximation}
To compare the proposed approximation with existing method, in this subsection, we derive the expression of beta approximation, which requires the first and the second moment of the network.

\begin{lemma}[Laplace Transform of the Interference]\label{lemma_laplace}
The Laplace transform of the interference is given by
\begin{align}
	\mathcal{L}_{I}(s) & = \exp\bigg(-2\pi\lambda\int_{0}^{\infty}\int_{0}^{r}\frac{2\pi\lambda\exp(-\lambda\pi r^2)}{1+(s p_{t}^{\prime}(x))^{-1} r^{\alpha} }{\rm d}x r{\rm d}r\bigg).
\end{align}
\end{lemma}
\begin{IEEEproof}
See Appendix \ref{app_laplace}.
\end{IEEEproof}

Next, we compute the $b$-th moment of the network.

\begin{lemma}[$b$-th Moment]\label{lemma_b_moment}
The $b$-th moment of the conditional success probability is
\begin{align}
	M_b(\theta) &= \int_{0}^{\infty}f_{R_u}(z)\exp\bigg(-\kappa_b(z) \sigma^2 \bigg)\mathcal{L}_{I}(\kappa_b(z)){\rm d}z,
\end{align}
where $\kappa_b(r) = b\frac{\theta r^{\alpha}}{ p_{t}(r)}$.
\end{lemma}
\begin{IEEEproof}
See Appendix \ref{app_b_moment}.
\end{IEEEproof}

Typically, (\ref{eq_meta_ext}) is solved by using Gil-Pelaez theorem \cite{gil1951note}, which requires the imaginary moments,
\begin{align}
\bar{F}_{P_s}(\theta,\gamma) = \frac{1}{2}+\frac{1}{\pi}\int_{0}^{\infty}\frac{1}{t}\Im(\exp(-jt\log\gamma)M_{jt}(\theta)){\rm d}t,\label{eq_meta_exactF}
\end{align}
where $j$ is the imaginary unit and $M_{jt}(\theta)$ is obtained by replacing the $b$ in the $b$-th moment by $jt$ and $\Im(\cdot)$ is the imaginary part of a complex number.

Generally, computing (\ref{eq_meta_exactF}) is tricky since it requires the imaginary moments. Alternatively, we can use beta approximation to compute, which only requires the first and the second moment and highly improves the analysis tractability, and this approximation shows good matching for a wide range of scenarios \cite{haenggi2015meta}.

\begin{remark}[Beta Approximation]
The beta approximation is given by
\begin{align}
	\bar{F}^{\prime\prime}_{P_s}(\theta,\gamma)	\approx 1- &I_{\gamma}\bigg(\frac{M_1(\theta)(M_1(\theta)-M_2(\theta))}{M_2(\theta)-M_1^2(\theta)},\nonumber\\
	& \frac{(M_1(\theta)-M_2(\theta))(1-M_1(\theta))}{M_2(\theta)-M_1^2(\theta)}
	\bigg),\label{eq_MetaBetaApp}
\end{align}
where $I_x(a,b)= \frac{\int_{0}^{x}t^{a-1}(1-t)^{b-1}{\rm d}t}{B(a,b)}$, and $
B(a,b) = \int_{0}^{1}t^{a-1}(1-t)^{b-1}{\rm d}t$. 	
\end{remark}

\section{Numerical Results}

\begin{figure*}[ht]
	\centering
	\subfigure[$\epsilon = 0.4$]{\includegraphics[width = 0.9\columnwidth]{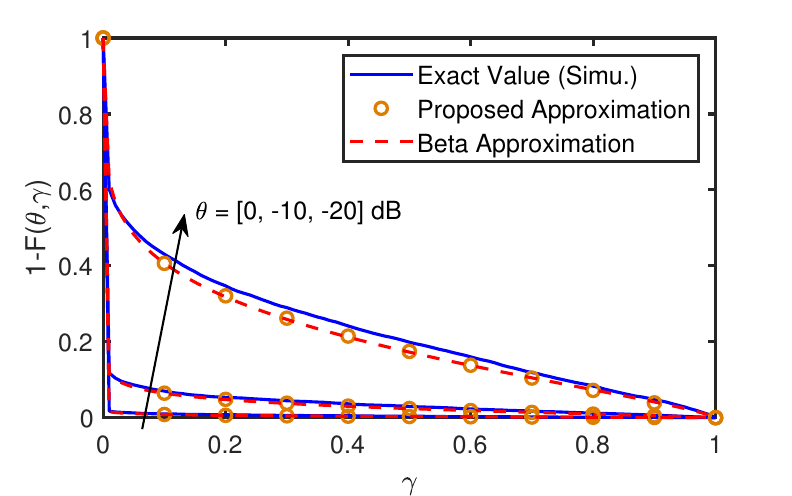}}
	\subfigure[$\epsilon = 0.4$]{\includegraphics[width = 0.9\columnwidth]{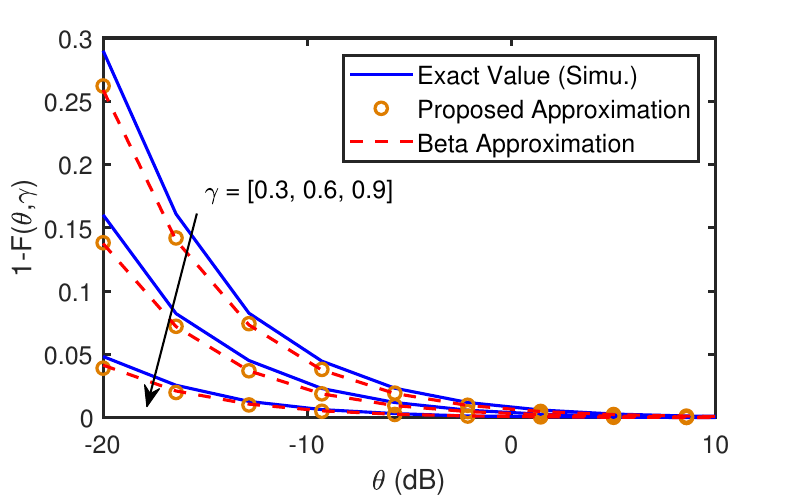}}
	\subfigure[$\epsilon = 0.8$]{\includegraphics[width = 0.9\columnwidth]{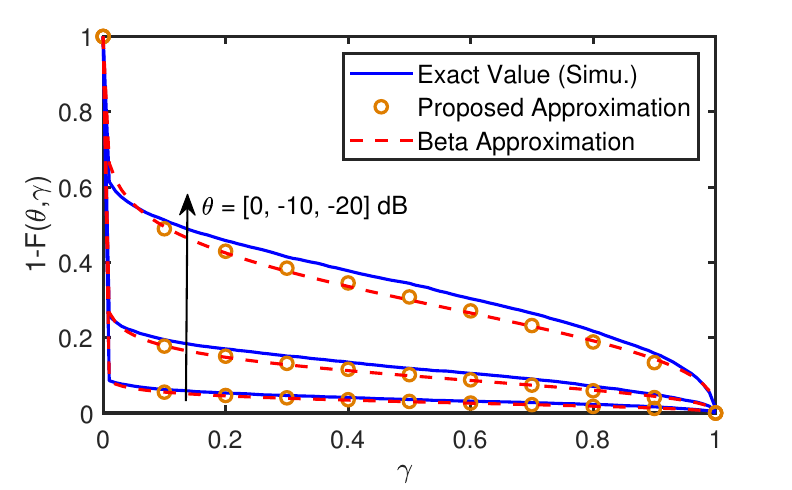}}
	\subfigure[$\epsilon = 0.8$]{\includegraphics[width = 0.9\columnwidth]{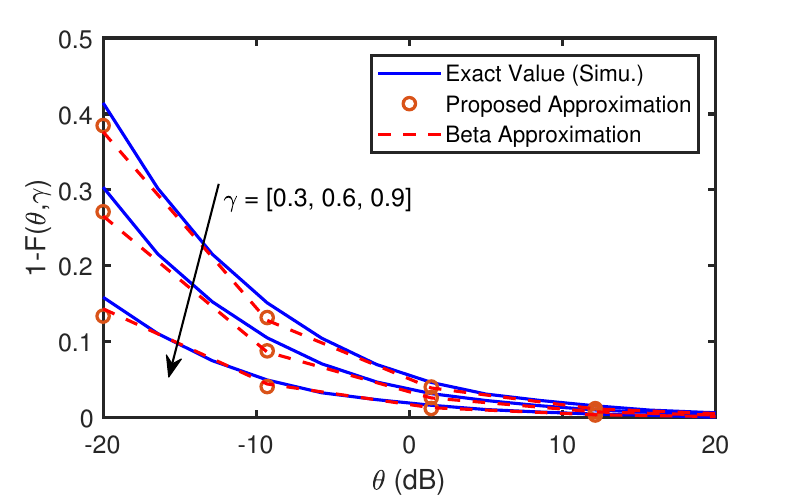}}
	\caption{ SINR meta distribution of uplink transmission of PPP networks. The solid lines are exact values based on simulations, dash lines are for beta approximation, and markers for the proposed approximation, respectively.
	}
	\label{fig_eta}
\end{figure*}
\begin{figure*}[ht]
	\centering
	\subfigure[]{\includegraphics[width = 0.9\columnwidth]{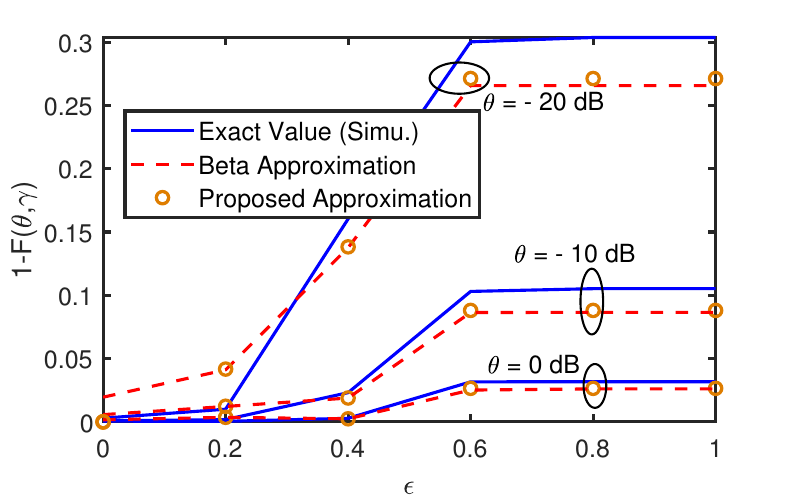}}
	\subfigure[]{\includegraphics[width = 0.9\columnwidth]{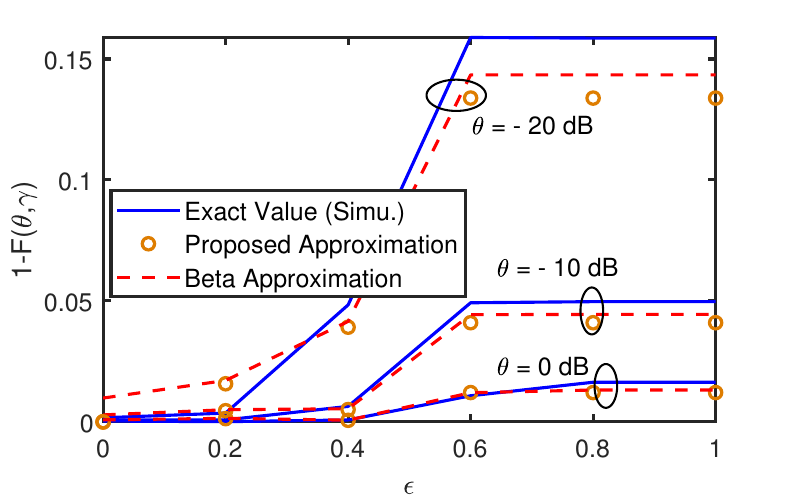}}
	\caption{ SINR meta distribution of uplink transmission of PPP networks at \textbf{(a)} $\gamma = 0.6$ and \textbf{(b)} $\gamma = 0.9$. The solid lines are exact values based on simulations, dash lines are for beta approximation, and markers for the proposed approximation, respectively.
	}
	\label{fig_gamma}
\end{figure*}

In this section, we validate the proposed approximation and compare it with the beta approximation and the exact value of the SINR meta distribution via Monte-Carlo simulations with a large number of iterations to ensure accuracy.
For the simulation of the considered setup, we first generated two realizations of PPPs for the locations of BSs and UE. UE is located in the Voronoi cells formed by BSs and in each resource block only one UE is active. Next, we compute the conditional success probability of each link (for each fixed realization, channel fading varies for $10^{4}$ times to obtain the conditional success probability) and obtain the CCDF of the conditional success probability. Finally, we repeat this process for $10^2$ times.
Unless stated otherwise, we use the system parameters listed herein. The maximum transmit power of UE is $p_{\rm max} = 200$ mW, the power control factor is $\rho = 0.008$ mW, the noise power is set as $\sigma^2 = 10^{-9}$ W, and the path-loss exponent is $\alpha = 4$.

In Fig. \ref{fig_eta}, we plot the exact value, beta approximation, and the proposed approximation of SINR meta distribution against $\theta$ or $\gamma$ at different values of $\gamma$ or $\theta$ under $\epsilon = 0.4,0.8$, respectively. The proposed approximation shows good matching at all values of $\gamma$ and $\theta$. Slightly different from the downlink scenario in \cite{10066317}, the results of the proposed approximation is very close to beta approximation, and difficult to tell which approximation has a better performance. For the time of computing one point, the proposed approximation is slightly shorter than the beta approximation. Even though the proposed approximation does not provide a dramatic improvement in performance, we still believe it is meaningful. While the beta approximation behaves more like a fitting of meta distribution, as mentioned in \cite{haenggi2015meta}, the reason for using beta distribution is that its high order moments fit well with the moments of the network, and some other approximations, such as nearest neighbor-only approximation, Markov and Chebyshev bounds, show gaps with SINR meta distribution, the proposed method provides a good matching approximation to the meta distribution. That is, in some special cases, e.g., Rayleigh fading models, (\ref{eq_meta_ext}) can be solved by using the dominant interferer-based approximation, and from this point of view, the proposed approximation completes the analysis of meta distribution.

Fig. \ref{fig_gamma} provides some insights about the effect of $\epsilon$ on the proposed approximation and SINR meta distribution. With the increases of $\epsilon$, the system reliability first increases slowly, then increases dramatically until the maximum achievable value which is limited by the maximum transmit power and interference. This is because of the transmit power of UE and distance to the serving BS. While the increase of $\epsilon$ increases the transmit power of UE and compensates for the path-loss, it also increases the interference. Therefore, an optimal $\epsilon$ exists to maximize the system's reliability. Besides, we notice that the proposed approximation, as well as the beta approximation, shows a small gap with the exact value of meta distribution at low values of $\theta$.

\section{Conclusion}
This work analyzes a dominant interferer-based approximation in uplink networks. Compared with the downlink scenario, the uplink analysis is more complex since the truncated power inversion control is used and the locations of interferers follow a non-homogeneous PPP. We first show that the proposed approximation is operable in an uplink scenario which follows a similar way as downlink scenarios. To validate the accuracy, we compare it with the traditional method, beta approximation, as well as the exact value obtained via simulation. The proposed approximation shows a good matching results in uplink scenarios for different values of the path-loss compensation factor. Consequently, this work completes the analysis of the dominant interferer-based approximation of meta distribution and proves that the proposed approximation works well in both uplink and downlink scenarios.

\appendix

\subsection{Proof of Lemma \ref{lemma_approx_int}}\label{app_approx_int}
Similar to \cite[Lemma 2]{10066317}, we consider the nearest interferer exactly and the rest interferers in mean sense. With that been said, conditioned on the location of the nearest interferer, we compute the mean of the rest of interferers. Consequently, the aggregate interference is approximated by the exact term of nearest interferer and the conditional mean of the rest of the terms,
\begin{align}
	I_1 &= \sum_{x\in\Phi_i} H_x p_{t,x}(R_{i,x}) D_{i,x}^{-\alpha}\nonumber\\
	&= H_{x_1} p_{t,x_1}(R_{i,x_1}) D_{i,x_1}^{-\alpha} + \sum_{x\in\Phi_i\setminus x_1} H_x p_{t,x}(R_{i,x}) D_{i,x}^{-\alpha}\nonumber\\
	&\approx H_{x_1} p_{t,x_1}(R_{i,x_1}) D_{i,x_1}^{-\alpha} +G(D_{i,x_1}),
\end{align}
in which
\begin{align}
	G(D_{i,x_1}) &= \mathbb{E}\bigg[\sum_{x\in\Phi_i\setminus x_1} H_x p_{t,x}(R_{i,x}) D_{i,x}^{-\alpha}\bigg] \nonumber\\
	&	\stackrel{(a)}{=} \mathbb{E}\bigg[\sum_{x\in\Phi_i\setminus x_1}  p_{t,x}(R_{i,x}) D_{i,x}^{-\alpha}\bigg]\nonumber\\
	&\stackrel{(b)}{=}2\pi\lambda\int_{D_{i,x_1}}^{\infty}\int_{0}^{z}(1-\exp(-\pi \lambda z^2))\nonumber\\
	&\quad\times\min(p_{\rm max},\rho r^{\alpha\eta}) f_{R_u}(r|z)(r) z^{-\alpha+1} {\rm d}r{\rm d}z,
\end{align}	
in which step (a) follows from the fact that all fading gains are i.i.d. distributed with mean of unity and step (b) follows from Campbell's theorem \cite{haenggi2012stochastic} with conversion from Cartesian to polar coordinates.
\vspace{- 0.5 cm}

\subsection{Proof of Lemma \ref{lemma_Ps_appro}} \label{app_Ps_appro}
	By using the approximated interference given in the previous lemma, the conditional success probability is approximated by,
\begin{small}
	\begin{align}
	&\mathbb{P}({\rm SINR} > \theta \mid R_{u}) = \mathbb{P}\bigg(\frac{H p_{t}(R_u) R_{u}^{-\alpha}}{\sigma^2+I_1} > \theta \mid R_{u}\bigg)\nonumber\\
	&=\mathbb{P}\bigg(H > \frac{\theta(\sigma^2+I_1)}{p_{t}(R_u) R_{u}^{-\alpha}} \mid R_{u}\bigg)\nonumber\\
	&\approx \mathbb{P}\bigg(H >  \theta(\sigma^2+H_{x_1} p_t(R_{i,1})D_1^{-\alpha}+G(D_1)) \frac{R_{u}^{ \alpha}}{p_{t}(R_u)}   \mid R_{u}\bigg)\nonumber\\
	&= \mathbb{E}\bigg[\exp\bigg(\theta(\sigma^2+H_{x_1} p_t(R_{i,1})D_1^{-\alpha}+G(D_1)) \frac{R_{u}^{ \alpha}}{p_{t}(R_u)}\bigg)\bigg]\nonumber\\
	&\stackrel{(a)}{=} \exp\bigg(\theta(\sigma^2+G(D_{1})) R_{u}^{\prime -1}\bigg) \frac{1}{1+\theta p_t(D_{1})D_{1}^{-\alpha} R_{u}^{\prime -1}},
\end{align}
\end{small}
where step (a) follows from the use of MGF of exponential distribution.
\vspace{- 0.5 cm}

\subsection{Proof of Theorem \ref{theorem_meta_app}}\label{app_theorem_meta_app}
Recall that SINR meta distribution is the CCDF of the conditional success probability, therefore, approximated by
\begin{small}
	\begin{align}
	&\bar{F}_{P_s}^{\prime}(\theta,\gamma) = \mathbb{P}(P_s(\theta)>\gamma) \nonumber\\
	&\approx \mathbb{P}\bigg(\exp\bigg(\theta(\sigma^2+G(D_1)) p_{t}^{-1}(R_u) R_{u}^{ \alpha}\bigg) \nonumber\\
	&\quad\times\frac{1}{1+\theta p_t(R_{i,1})D_{1}^{-\alpha} p_{t}^{-1}(R_u) R_{u}^{ \alpha}}>\gamma\bigg)\nonumber\\
	&\stackrel{(a)}{=} \mathbb{E}\bigg[\mathbb{P}\bigg(p_{t}(R_u)R_{u}^{-\alpha}<-\frac{1}{\theta p_t(R_{i,1})D_{1}^{-\alpha} }\nonumber\\
	&\quad+\frac{1}{S(R_{1})}W\bigg(0,\frac{S(R_{i,1})\exp(S(D_{1})\theta^{-1}p_{t}^{-1}(R_{i,1})D_{1}^{ \alpha})}{\gamma\theta p_t(R_{i,1})R_{1}^{-\alpha}}\bigg)\bigg)\bigg]\nonumber\\
	&= \int_{0}^{\infty}F_{R_{u}^{\prime}}(K(r,\theta,\gamma))f_{D_1}(r){\rm d}r,
\end{align}
\end{small}
in which step (a) can be solved by using simple algebraic manipulations and the definition of the Lambert W function: $W(0,x)\exp(W(0,x)) = x$, and
\begin{small}
	\begin{align}
		&K(r,\theta,\gamma) = -\frac{1}{\theta p_t(R_{i,x_1})r^{-\alpha} } \nonumber\\
		&\quad+\frac{1}{S(r)}W\bigg(0,\frac{S(r)\exp(S(r)\theta^{-1}p_{t}^{-1}(R_{i,1})r^{ \alpha})}{\gamma\theta p_t(R_{i,1})r^{-\alpha}}\bigg) \nonumber\\
		&\stackrel{(a)}{\approx} -\frac{1}{\theta \bar{p}_{t}(r) r^{-\alpha} } +\frac{1}{S(r)}W\bigg(0,\frac{S(r)\exp(S(r)\theta^{-1}\bar{p}_{t}^{-1}(r)r^{ \alpha})}{\gamma\theta \bar{p}_{t}(r) r^{-\alpha}}\bigg),
	\end{align}
\end{small}
in which step (a) is approximated by taking the integration of $p_{t}(r)$ separately.

Finally,  the CDF of $R_{u}^{\prime}$ is computed by
\begin{align}
	&F_{R_{u}^{\prime}}(x) = \mathbb{P}(R_{u}^{\prime}<x) = \mathbb{P}(p_{t}^{\prime}(R_u) R_{u}^{-\alpha}<x) \nonumber\\
	&= \left\{
	\begin{aligned}
		&\mathbb{P}(\rho R_{u}^{(1-\alpha)\epsilon}<x),\quad \text{if}\quad 0< R_u < \bigg(\frac{p_{\max}}{\rho}\bigg)^{\frac{1}{\alpha\epsilon}},\\
		&\mathbb{P}(p_{\rm max} R_{u}^{-\alpha}<x), \quad \text{if}\quad R_u \geq \bigg(\frac{p_{\max}}{\rho}\bigg)^{\frac{1}{\alpha\epsilon}},\\
	\end{aligned}
	\right.
\end{align}
proof completes by using  $F_{R_{u}}(r)$.

\subsection{Proof of Lemma \ref{lemma_laplace}}\label{app_laplace}
The Laplace transform of the aggregate interference is given by
\begin{align}
	&\mathcal{L}_{I}(s) = \mathbb{E}_{I}[\exp(-sI)] \nonumber\\
	&= \mathbb{E}_{I}\bigg[\exp\bigg(-s\sum_{x\in\Phi_i} H_x p_{t,x}(R_{i,x}) D_{i,x}^{-\alpha}\bigg)\bigg]\nonumber\\
	&= \mathbb{E}_{I}\bigg[\prod_{x\in\Phi_i} \exp\bigg(-s H_x p_{t,x}(R_{i,x}) D_{i,x}^{-\alpha}\bigg)\bigg] \nonumber\\
	&= \mathbb{E}_{I}\bigg[\prod_{x\in\Phi_i} \mathbb{E}\bigg[\frac{1}{1+s p_{t,x}(R_{i,x}) D_{i,x}^{-\alpha} }\bigg]\bigg]\nonumber\\
	&= \mathbb{E}_{I}\bigg[\prod_{x\in\Phi_i} \int_{0}^{D_i}\frac{1}{1+s p_{t}^{\prime}(x) D_{i,x}^{-\alpha} }f_{R_i}(x|D_i){\rm d}x\bigg] \nonumber\\
	&= \exp\bigg(-2\pi\lambda\int_{0}^{\infty}(1-\exp(-\pi\lambda r^2))\nonumber\\
	&\times\bigg(1- \int_{0}^{r}\frac{1}{1+s p_{t}^{\prime}(x) r^{-\alpha} }f_{R_i}(x|r){\rm d}x\bigg)r{\rm d}r\bigg) \nonumber\\
	&= \exp\bigg(-2\pi\lambda\int_{0}^{\infty}\int_{0}^{r}\frac{2\pi\lambda\exp(-\lambda\pi r^2)}{1+(s p_{t}^{\prime}(x))^{-1} r^{\alpha} }{\rm d}x r{\rm d}r\bigg) 
\end{align}	

\subsection{Proof of Lemma \ref{lemma_b_moment}}\label{app_b_moment}
	The $b$-th moment is computed by taking the expectation of the $b$-th moment of the conditional success probability,
\begin{align}
	M_b(\theta) &= \mathbb{E}[P_{s}^b(\theta)] ,
\end{align}	
in which the exact expression of the conditional success probability (different from (Lemma \ref{lemma_Ps_appro})) is given by
\begin{align}
	P_s(\theta) &= \mathbb{P}({\rm SINR}>\theta|R_u) = \mathbb{P}\bigg( \frac{H p_{t}(R_u) R_{u}^{-\alpha}}{\sigma^2+I} >\theta|R_u\bigg) \nonumber\\
	&= \mathbb{P}\bigg( H >\frac{\theta R_{u}^{\alpha}}{ p_{t}(R_u)}(\sigma^2+I)|R_u\bigg) \nonumber\\
	&= \mathbb{E}_{I}\bigg[\exp\bigg(-\frac{\theta R_{u}^{\alpha}}{ p_{t}(R_u)}(\sigma^2+I)\bigg)\bigg] \nonumber\\
	&= \exp\bigg(-\frac{\theta R_{u}^{\alpha}}{ p_{t}(R_u)} \sigma^2 \bigg)\mathcal{L}_{I}\bigg(\frac{\theta R_{u}^{\alpha}}{ p_{t}(R_u)}\bigg).
\end{align}
\bibliographystyle{IEEEtran}
\bibliography{Ref18}
\end{document}